\documentclass[10pt, conference, letterpaper]{IEEEtran}

\IEEEoverridecommandlockouts                              % This command is only
\usepackage{amsfonts}    % Enable this line and disable the
\usepackage{color}
\usepackage{graphicx}
\usepackage[dvips]{epsfig}
\usepackage{graphics}
\usepackage{arydshln}
\usepackage{amsmath} % assumes amsmath package installed
\usepackage{amssymb}  % assumes amsmath package installed
\usepackage{subfigure}
\usepackage{subfloat}
\usepackage{caption}
\usepackage{algorithm}
\usepackage{algorithmic}
\usepackage{hyperxmp}
\usepackage{hyperref}

\def\bZ{\mathbb{Z}}

\def\cA{\mathcal{A}}
\def\cB{\mathcal{B}}

\def\cH{\mathcal{H}}

\def\cN{\mathcal{N}}

\def\bee{\begin{equation}}
\def\ene{\end{equation}}
\def\beq{\begin{eqnarray}}
\def\enq{\end{eqnarray}}
\def\been{\begin{equation*}}
\def\enen{\end{equation*}}

\newtheorem{ass}{Assumption}[section]
\newtheorem{pro}{Proposition}[section]

\newtheorem{thm}{Theorem}[section]

\begin{document}

\title{Dynamic Pricing and Capacity Allocation of UAV-provided Mobile Services}

\author{Xuehe~Wang~and~Lingjie~Duan%,~\IEEEmembership{Member,~IEEE}% <-this % stops a space
\thanks{X. Wang and L. Duan are with the Pillar of Engineering Systems and Design, Singapore University of Technology and Design, Singapore 487372 (e-mail: xuehe\_wang@sutd.edu.sg; lingjie\_duan@sutd.edu.sg).}}

\maketitle

\begin{abstract}

Due to its agility and mobility, the unmanned aerial vehicle (UAV) is a promising technology to provide high-quality mobile services (e.g., fast Internet access, edge computing, and local caching) to ground users. Major Internet Service Providers (ISPs) want to enable UAV-provided services (UPS) to improve and enrich the current mobile services for additional profit. This profit-maximization problem is not easy as the UAV has limited energy storage and needs to fly closely to serve users, requiring an optimal energy allocation for balancing both hovering time and service capacity. When hovering in a hotspot, how the UAV should dynamically price its capacity-limited UPS according to randomly arriving users with private service valuations is another question. We prove that the UAV should ask for a higher price if the leftover hovering time is longer or its service capacity is smaller, and its expected profit approaches to that under complete user information if the hovering time is sufficiently large. As the hotspot's user occurrence rate increases, a shorter hovering time or a larger service capacity should be allocated. Finally, when the UAV faces multiple hotspot candidates with different user occurrence rates and flying distances, we prove that it is optimal to deploy the UAV to serve a single hotspot. With multiple UAVs, however, this result can be reversed with UAVs' forking deployment to different hotspots.

\end{abstract}

\section{Introduction}

As the demand for mobile services increases exponentially, it is imperative for the Internet Service Providers (ISPs) to improve existing services' capacity and coverage and provide more customized services for profit maximizing. Due to its agility and mobility, the unmanned aerial vehicle (UAV) emerges as a promising vehicular technology to provide value-added mobile services (e.g., fast Internet access, edge computing, and local caching) to ground users. For example, AT\&T has designed a Flying COW to beam high-throughput wireless coverage to the crowds in sports stadiums \cite{droneatt}. By endowing with edge computing capability, the UAV can be also used to offer computation offloading services to mobile users with limited terminal processing capability \cite{jeong2018mobile}. The cache-enabled UAV is also implemented recently to improve the quality-of-experience of mobile devices by caching and distributing the popular content to them \cite{chen2017caching}. Major ISPs want to enable such UAV-provided services (UPS) to improve and enrich their mobile services for additional profit. For example, Verizon hired a specialized drone company Skyward to provide compatible value-added services to its users \cite{verizonskyward}. The global revenue of UPS is expected to increase from \$792 million in 2017 to \$12.6 billion by 2025 \cite{UAVrevenue}.

%(e.g., congested WiFi)

%help developers and businesses better create and manage drones that also happen to
The literature focuses on the technological issues of enabling UPS such as exploiting air-to-ground communication to enlarge wireless coverage and addressing UAV energy constraints. For example, \cite{mozaffari2015drone} analyzes the optimal operating altitude for a UAV's maximum wireless coverage by considering the trade-off between the opportunity of line-of-sight transmission and signal attenuation. In \cite{di2015energy}, an energy-aware UAV path planning algorithm is proposed to minimize energy consumption of covering users in a specific area. In the UAV-provided edge computing application, \cite{jeong2018mobile} studies the optimization of the UAV's trajectory and computing offloading under its energy constraint. As for local caching application, in \cite{chen2017caching} the optimal UAV's location and the content to cache are jointly investigated according to the users' content request distribution and their mobility patterns. In \cite{zhang2018fast}, two fast UAV deployment problems are studied for optimal wireless coverage. \cite{zhang2018economics}, \cite{xu2018uav} further study the UAV placement games for best serving all the users while ensuring all selfish users' truthfulness in reporting their locations. \cite{WangZhe2018Globecom} focuses on adapting the UAV deployment to the spatial randomness of mobile users in order to maximize the average throughput of all users in the uplink information transmission.

\begin{figure}
\centering\includegraphics[scale=0.33]{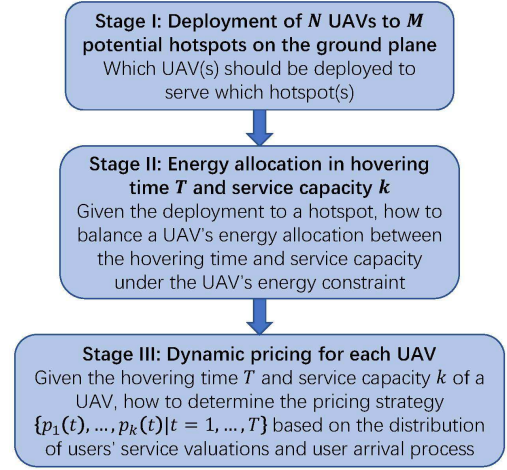}\caption{Three-stage UPS provision of the UAV firm for profit-maximization.}\label{fig_model3stage}
\end{figure}

Still, the economic issues of UPS for serving mobile users are largely overlooked in the literature, hindering the successful development of UPS in the long run. As the UAV's hovering in the air and its service (e.g., computing and caching) provision to ground users are both energy-consuming, a longer hovering time helps meet more demands yet leaving less energy for servicing them. How to balance the hovering time and service capacity under the limited energy budget is critical to ensure the economic viability of UPS. Further, when hovering in a hotspot for a given time period, how to dynamically price the capacity-limited UPS to ground users for profit-maximization is another issue. This is challenging under incomplete information about the mobile users' randomness in arriving and their private valuations of buying UPS. What's more, when facing multiple hotspot candidates with different user occurrence rates and flying distances, the optimal deployment of multiple UAVs to cooperatively serve the chosen hotspots needs to be studied. This paper proposes a three-stage UPS provision model to study these economic issues as shown in Fig. \ref{fig_model3stage}: first on multiple UAVs' deployment to cooperatively cover heterogeneous hotspots, then on energy allocation of each UAV to balance hovering time and service capacity for its chosen hotspot to deploy, and finally the dynamic UPS pricing for each UAV over its hovering time. These three stages following time sequence are inter-dependent for maximizing the UPS profit and we will apply backward induction for analyzing them.

It should be noted that in the literature there are some related works studying the pricing issues for resource constrained wireless networks (e.g., \cite{gizelis2011survey}, \cite{niyato2007wireless}, \cite{musacchio2006wifi}, \cite{duan2011investment}). For example, \cite{gizelis2011survey} discusses both the static and dynamic pricing schemes for a wireless network, depending on whether the network can adapt to the service requirements of their subscribers. An optimal pricing for bandwidth sharing is proposed in \cite{niyato2007wireless} for an integrated network using different wireless technologies. \cite{musacchio2006wifi} studies dynamic WiFi pricing for a fixed user over time under incomplete information of the user's service valuation and utilities. In the broader literature of economics and operations research, there are also some related works about the dynamic admission control and pricing of generic services for users under incomplete demand information (\cite{gershkov2014dynamic},\cite{stadje1990full},\cite{albright1974optimal}). However, these works assume a fixed service capacity and do not consider users' randomness in arrivals, while in this paper studies a more difficult scenario that each UAV has interchangeable energy capacities for hovering and servicing to further balance in practice and the mobile users are also randomly moving on the ground.

Our key novelty and main contributions are summarized as follows.
\begin{itemize}
  \item \emph{Economics of UAV-provided mobile services:} To our best knowledge, this paper is the first work studying the economics of UAV-provided services (UPS) provision, including the interdependent UAV-network deployment, capacity allocation and finally dynamic service pricing for UPS profit maximization in Fig. \ref{fig_model3stage}. By applying backward induction, we first analyze each UAV's dynamic pricing under user incomplete information for a given hovering time at a given hotspot deployment, then the optimal trade-off between its hovering time and service capacity at a given hotspot, and finally deployment of multiple cooperative UAVs to cover heterogeneous hotspots.
  \item \emph{Dynamic UPS pricing under incomplete information (Section \ref{sec_pricing}):} Under the incomplete information regarding users' random arrivals and private service valuations, we propose an optimal dynamic pricing scheme for each UAV and prove that the UAV should ask for a higher UPS price if its leftover hovering time is longer or its service capacity is smaller. Its expected profit approaches to that under complete user information if the hovering time is sufficiently large. We also show that the UAV may take advantage of a large variance of users' valuation distribution.
  \item \emph{Optimal energy allocation to balance hovering time and service capacity (Section \ref{sec_energyallocation}):} Though a longer hovering time ensures a higher service price under incomplete information, it leaves a smaller service capacity under the total energy budget, the UAV should balance its hovering time and service capacity to maximize its profit for providing UPS. To characterize the tradeoff, we develop an optimal threshold-based capacity allocation policy which is easy to implement. We show that as the hotspot's user occurrence rate increases, a smaller hovering time or a larger service capacity should be allocated.
  \item \emph{Cooperative UAVs' deployment to heterogeneous hotspots (Section \ref{sec_UAVdeployment}):} We study the deployment of multiple cooperative UAVs owned by a UAV company to heterogeneous potential hotspots for maximizing the total UPS profit. A hotspot's high user occurrence rate helps ensure a large demand for UPS yet its flying distance should not be far for a UAV to reach under energy constraint. We aim to reach the best trade-off between different hotspots' user occurrence rates and flying distances for UAV deployment. We prove that it is optimal for a single UAV to only serve one hotspot. However, when we have multiple UAVs for deployment, they should fork to serve different hotspots especially when hotspots are more symmetric or the UAV number is large. When multiple UAVs are deployed to the same hotspot, they can cooperatively pool their service capacities yet waste more energy in hovering at the same time.
\end{itemize}

\section{System Model and Problem Formulation}\label{sec_model}

\begin{figure}
\centering\includegraphics[scale=0.38]{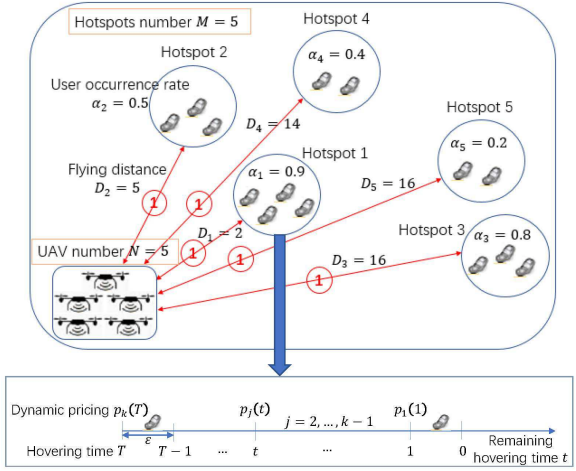}\caption{An example of $N=5$ UAVs' deployment to $M=5$ potential hotspots with different user occurrence rates $\alpha_m$'s and flying distances $D_m$'s, followed by the dynamic pricing with service capacity $k$ in the hovering period $T$ for each UAV at its hotspot. The number in red circle indicates how many UAVs are assigned to the corresponding hotspot.}\label{fig_modelUAVs}
\end{figure}

For maximizing the UAV company's profit, we propose a three-stage UPS provision model to study the UAVs' optimal deployment, capacity allocation and dynamic pricing as shown in Fig. \ref{fig_model3stage}. In Stage $I$, we deploy a number $N$ of identical UAVs from the UAV station to $M$ potential heterogeneous hotspots to cooperatively serve users there.\footnote{In practice, most users are clustered in hotspots (e.g., shopping malls and residential areas). Actually, for any unpopular place, we can still model it as a hotspot here, by updating its low user occurrence rate and its flying distance for the UAV to reach.} Fig. \ref{fig_modelUAVs} shows an example of deploying $N=5$ UAVs to $M=5$ heterogeneous hotspots. Each hotspot $m$'s user occurrence rate and flying distance from the UAV station are denoted as $\alpha_m$ and $D_m, m=1,...,M$, respectively. Users' random arrivals at a hotspot in a discrete time horizon and each time slot's duration are properly selected such that there is at most one user occurrence at a time. Here $\alpha_m$ also tells the probability of having a user's occurrence in each time slot at hotspot $m$. Each user has a one-time-slot service session with the UAV and then leaves the hotspot.\footnote{Our analysis can also be extended to the case that a user stays in service for more than one time slot.}%\footnote{The time slots are partitioned such that there is at most one user appearing in each time slot. If there are more than one user appearing in a time slot, we can further partition the time period into smaller time slots.}

In Stage $II$, given an individual UAV's energy storage $B$ upon arrival at its deployed hotspot, this UAV should decide the energy allocation to hovering time $T$ and service capacity $k$ with $T+ck\leq B$, where $c>0$ is the energy consumption for serving a user (in relative sense to the energy consumption per unit hovering time). Both $T$ and $k$ are integers, telling how many time slots for hovering and how many users to serve, respectively. If the UAV hovers longer, it may encounter more users and charge higher prices under user incomplete information, yet the final number $k$ of users it can serve decreases given the total energy budget $B$.

In Stage $III$, given hovering time $T$ and service capacity $k$ for this hotspot,
our objective is to maximize the UAV's expected profit $R_k(T)$ at this hotspot by designing the dynamic pricing $\{p_1(t),...,p_k(t)|t=1,...,T\}$ in the discrete time horizon as shown in the lower part of Fig. \ref{fig_modelUAVs}, where
$p_j(t),j=1,...,k$, is the price for selling the UPS to the $j$th-to-last user at $t$ time units before the end of hovering/selling period $T$. Note that $t=0$ (or $t=T$) is the end (beginning) of the hovering interval and $p_1(t)$ (or $p_k(t)$) is the price for serving the last (first) user. For simplicity, we will use ``$t$ time units before the end of hovering period $T$" and ``time $t$" interchangeably in the rest of the paper. A user (if occurs) will accept the price if his service valuation $v$ is greater than the price asked by the UAV. It is assumed that there is no possibility to recall the users and they may already leave. The users' valuations of the UPS are independent and identically distributed (i.i.d.) according to a probability density function (PDF) $f(v), v\in[a,b]$. Though all potential users' valuations follow the same distribution, their realized valuations are different in general. Under the incomplete information, the UAV does not know the user occurrence for UPS over time $t$ or the user's private service valuation $v$. It only knows the user occurrence probability in each time slot and the valuation distribution $f(v)$.

For economic purpose, we want to maximize the total UAVs' final profit in the three-stage UPS provision model under their energy budgets. In the following, we will use backward induction to first analyze the optimal dynamic pricing for each UAV in Stage $III$ with given service capacity and hovering time at a given hotspot, then the energy allocation for this UAV at the given hotspot in Stage $II$, and finally the UAVs' optimal deployment to all possible hotspots in Stage $I$. %Due to the page limitation, some of the proofs are given in the technical report \cite{Wang_tech}.

\section{Dynamic UPS pricing under Incomplete Information in Stage III}\label{sec_pricing}% under Incomplete Information

%Mobile users are moving and their service requests in every time slot are unknown at the hostpot.

In Stage $III$, each UAV's hovering time $T$ and service capacity $k$ are given for its deployed hotspot $m$. The UAV at this hotspot has probability $\alpha_m$ or simply $\alpha$ of meeting a user request in each time slot. Here we skip the subscript as the analysis holds for a UAV at any hotspot. The UAV should decide the dynamic pricing $p_j(t)$ at $t$ time slots before the end of hovering interval $T$ given any leftover service capacity $j,j=1,...,k$ as shown in Fig. \ref{fig_modelUAVs}. Note that fixed pricing rate is not optimal and is just a special case of our dynamic pricing here. It is possible to deploy more than one UAV at the same hotspot and their cooperative pricing is studied later in Section \ref{sec_UAVdeployment}. %\footnote{We will extend the pricing analysis for more than one UAV at the same hotspot in Section \ref{sec_UAVdeployment}.}

Before studying the UAV's optimal pricing strategy for any service capacity $k$, we first consider the case of $k=1$. That is, the UAV can only serve one user finally. By announcing price $p_1(t)$ at time $t$, a user (if appears with probability $\alpha$) will accept and pay the price if his service valuation $v$ is greater, i.e., $v\geq p_1(t)$. Given the cumulative distribution function (CDF) of his service valuation $F(v)$, the probability that a user will appear and accept the price is $\alpha(1-F(p_1(t)))$. Then, the UAV's expected total profit in the remaining $t$ time slots is%Given the service is not sold during the previous time interval
\bee\label{equ_R1(t)}\begin{split} R_1(t)=&\alpha p_1(t)(1-F(p_1(t)))\\
&+R_1(t-1)(1-\alpha(1-F(p_1(t)))).\end{split}\ene

We then consider the case of $k=2$. After successfully serving the first user and using up one service capacity, the profit analysis of the case $k=1$ in (\ref{equ_R1(t)}) can be applied for subsequently serving the second user. Note that the expected total profit received from the second user at $t$ time slots is $R_1(t)$. By successfully serving the first user at price $p_2(t)$ at time $t$, the UAV's total profit at time $t$ is $p_2(t)+R_{1}(t-1)$. Otherwise, the UAV will keep $R_2(t-1)$ with two service quota for the remaining $t-1$ time slots. Then, the expected profit with service capacity $k=2$ at time slot $t$ is
\bee\begin{split} R_2(t)=&\alpha (p_2(t)+R_{1}(t-1))(1-F(p_2(t)))\\&+R_{2}(t-1)(1-\alpha(1-F(p_2(t)))).\end{split}\ene

Similar to the above analysis, for the general $k\geq 2$ case, the expected total profit at time $t$ can be derived recursively as
\bee\label{equ_R_k(t)}\begin{split} R_k(t)=&\alpha (p_k(t)+R_{k-1}(t-1))(1-F(p_k(t)))\\
&+R_{k}(t-1)(1-\alpha(1-F(p_k(t)))),\end{split}\ene
where $R_k(0)=0$ due to zero remaining hovering time and $R_0(t)=0$ due to zero leftover service capacity. %Obviously, $R_k(t)\geq R_{k-1}(t)$.

By taking the derivative of $R_k(t)$ with respect to $p_k(t)$, the optimal price $p_k(t)$ satisfies
\bee\begin{split}\label{equ_p_k(t)} \frac{dR_k(t)}{dp_k(t)}=&\alpha\Big(1-F(p_k(t))-f(p_k(t))\\
&\times\big(p_k(t)-(R_k(t-1)-R_{k-1}(t-1))\big)\Big)=0. \end{split}\ene
According to (\ref{equ_p_k(t)}), it is easy to check that $p_k(t)\geq R_k(t-1)-R_{k-1}(t-1)$. We can also derive the following result from (\ref{equ_p_k(t)}).

\begin{pro}\label{pro_Rktrelationship} For $\forall t\in\{1,\cdots, T\}$ and $k\geq 1$, we have $R_k(t)\geq R_k(t-1)$. Moreover, if $t<k$, $R_k(t)=R_{t}(t)$, and otherwise $kR_1(\lfloor\frac{t}{k}\rfloor)\leq R_k(t)$.
\end{pro}

Intuitively, if the remaining hovering time $t$ is less than the maximum number of users $k$ that the UAV can serve, i.e., $t<k$, the UAV can at most serve $t$ users (one per each time slot) and the extra service capacity is wasted, thus $R_k(t)=R_{t}(t)$ and the price $p_k(t)$ at time $t<k$ does not exist. If the UAV has reasonable leftover time $t$ to sell $k$ (i.e., $t\geq k$), it is better to optimize its expected profit $R_k(t)$ via jointly pricing $k$ service capacities over $t$ time slots, rather than independently pricing for each service capacity with separate selling time $\lfloor\frac{t}{k}\rfloor$. This implies that $R_k(t)$ concavely increases with $k$ and $t$.

In the following, we assume the distributions of the users' service valuations are regular, which is widely adopted in the realm of mechanism design \cite{ewerhart2013regular}.
\begin{ass} $\phi(v)=v-\frac{1-F(v)}{f(v)}$ is an increasing function of $v$, where $F(v)$ and $f(v)$ are the CDF and PDF of each user's valuation $v$, respectively.
\end{ass}

Regularity holds for many distributions such as uniform, normal, exponential and Rayleigh distributions.

\begin{pro}\label{pro_priceincrease} For $\forall t\in\{1,\cdots, T\}$ and $k\geq 1$, Algorithm \ref{alg1} optimally returns the dynamic pricing scheme with computation complexity $O(kT)$. Especially, when $k=1$, the optimal price $p_1(t)$ is a non-decreasing function of leftover time $t$ and mean user occurrence rate $\alpha$ in the hotspot.
\end{pro}

The proof of Proposition \ref{pro_priceincrease} is given in Appendix \ref{app_pro_priceincrease}.

\begin{algorithm}[t]
\caption{UAV's dynamic pricing for serving $k$ users in hovering time $T$}
\begin{algorithmic}[1]

%\STATE \textbf{Input:}
%Hover time $T$, service capacity $k$, hotspot's user occurrence rate $\alpha$, and CDF of each user's service valuation $F(v)$
%
%\STATE \textbf{Output:}
%Optimal pricing strategy $\{p_1(t),...,p_k(t)|t=1,\cdots,T\}$, and the expected profit $R_k(T)$

\FOR {$j = 1$ to $k$ }
%\STATE $R_j(0)=0$
\FOR {$t = 1$ to $T$ }
\IF {$t<=j-1$}
	\STATE {$R_j(t)=R_{t}(t)$;}
    %\STATE {$p_j(t)=0$;}
\ELSE
	\STATE {compute $p_j(t)$ as the unique solution to (\ref{equ_p_k(t)});}
	\STATE {update $R_j(t)$ according to $p_j(t)$ and (\ref{equ_R_k(t)});}
\ENDIF
\RETURN $p_j(t)$, $R_j(t)$\\
\ENDFOR
\ENDFOR
\RETURN $\{p_1(t),...,p_k(t)|t=1,\cdots,T\}$ and $R_k(T)$\\

\end{algorithmic}
\label{alg1}
\end{algorithm}

\begin{figure}
\centering\includegraphics[scale=0.332]{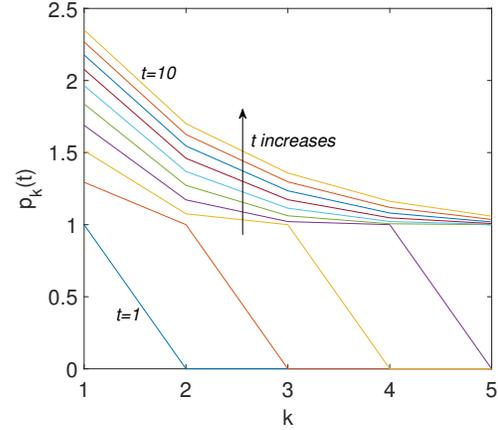}\caption{Optimal price $p_k(t)$ at each time $t$ versus the service capacity $k$ for exponential distributions of users' service valuations when $\alpha=0.8$, $\lambda=1$ and $T=10$.}\label{fig_pricek}
\end{figure}

In Algorithm \ref{alg1}, we first compute the optimal price $p_j(t)$ according to (\ref{equ_p_k(t)}), which is a function of $R_{j-1}(t-1)$ and $R_j(t-1)$ with initial conditions $R_{1}(0)=0$ and $R_0(t)=0$. Note that the price $p_j(t)$ at time $t\leq j-1$ does not exist as the leftover time slots $t$ is not enough to serve the $j$th users. Finally, we can obtain the expected profit $R_j(t)$ based on $p_j(t)$. Proposition \ref{pro_priceincrease} also shows that the UAV should ask for a higher price if it has more leftover time $t$ for encountering more users or a larger user demand (characterized by a higher user occurrence rate $\alpha$). Actually, for general $k$, we can numerically show that the optimal price $p_k(t)$ has the following properties (see Fig. \ref{fig_pricek}):
\begin{itemize}
  \item For any given $t$, the optimal price $p_k(t)$ decreases with $k$, as the UAV has more service capacity supply $k$ to meet the users' demand.
  \item For any given $k$, the optimal price $p_k(t)$ increases with $t$, as the UAV has more leftover time for encountering more users.
\end{itemize}

\subsection{Continuous-time Relaxation For More Tractable Analysis}\label{sec_continuous_pricing}

In the discrete time model as in Fig. \ref{fig_modelUAVs}, we can only use a recursive and numerical way to derive $R_k(t)$ and $p_k(t)$ according to (\ref{equ_R_k(t)}) and (\ref{equ_p_k(t)}). To obtain more analytical results for dynamic pricing design, we next apply continuous-time relaxation on the discrete time model. Assume users arrive according to a Poisson process with arrival rate $\alpha'$. Denote the time duration of each time slot for the discrete time model as $\varepsilon$ in Fig. \ref{fig_modelUAVs}.
To keep the same user occurrence rate $\alpha$ per time slot as in the discrete time case, we have $\varepsilon\alpha'+o(\varepsilon)=\alpha$ as $\varepsilon\rightarrow 0$. Similar to the analysis of the discrete time case in (\ref{equ_R_k(t)}), as $\varepsilon\rightarrow 0$, the expected total profit that the UAV can obtain at continuous time $t+\varepsilon$ is
\bee\label{equ_expo_continuous_derive}\begin{split} R_k(t+\varepsilon)&=\alpha'\int_t^{t+\varepsilon}(p_k(x)+R_{k-1}(x))(1-F(p_k(x)))dx\\
&+R_k(t)(1-\alpha'\int_t^{t+\varepsilon}(1-F(p_k(x)))dx)+o(\varepsilon).\end{split}\ene
Note that $R_k(0)=0$. According to (\ref{equ_expo_continuous_derive}), the UAV's expected total profit with service capacity $k$ at time $t$ can be derived as
\bee\label{equ_Rk_continuous_derive} R_k(t)=\alpha'\int_0^t(p_k(x)+R_{k-1}(x)-R_k(x))(1-F(p_k(x)))dx. \ene
To ensure positive profit, $p_k(x)\geq R_k(x)-R_{k-1}(x)$. The optimal price $p_k(t)$ that maximizes the expected profit $R_k(t)$ is simplified to
\bee\label{equ_solve_sup_pk} p_k(t)=\arg\max_{p\geq R_{k}(t)-R_{k-1}(t)}(p+R_{k-1}(t)-R_k(t))(1-F(p)). \ene

%For other distributions of service valuations, it is difficult to obtain the closed-form expected profit.
To analytically obtain the expected profit, we further consider the case that the users' i.i.d. service valuations follow exponential distributions, i.e., $F(v)=1-e^{-\lambda v}$.\footnote{The analysis method also holds for other continuous distributions though the analysis is move involved without closed-form.} Then, by solving (\ref{equ_solve_sup_pk}), the optimal price $p_k(t)$ is \bee\label{equ_pk_expo_continuous} p_k(t)=\frac{1}{\lambda}+R_k(t)-R_{k-1}(t), \ene
which is greater than the mean valuation $\frac{1}{\lambda}$ by considering the future pricing opportunity characterized by $R_k(t)-R_{k-1}(t)\geq 0$. Insert (\ref{equ_pk_expo_continuous}) into (\ref{equ_Rk_continuous_derive}), the expected profit for serving $k$ users in total hovering time $T$ can be derived in closed-form, given by
\bee\label{equ_Rk_expo_go0} R_k(T)=\frac{1}{\lambda}\log\bigg(\sum_{i=0}^k\frac{1}{i!}(\frac{\alpha' T}{e})^i\bigg). \ene
From (\ref{equ_pk_expo_continuous}) and (\ref{equ_Rk_expo_go0}), we can obtain the closed-form dynamic price at time $t\in[0,T]$ explicitly as
\bee\label{equ_pk_expo} p_k(t)=\frac{1}{\lambda}+\frac{1}{\lambda}\log\bigg(\frac{\sum_{i=0}^k\frac{1}{i!}(\frac{\alpha' t}{e})^i}{\sum_{i=0}^{k-1}\frac{1}{i!}(\frac{\alpha' t}{e})^i}\bigg). \ene

\begin{pro}\label{pro_pk_expo_go0} The optimal expected profit $R_k(T)$ in (\ref{equ_Rk_expo_go0}) concavely increases with both $k$ and $T$, respectively. Further, the optimal price $p_k(t)$ in (\ref{equ_pk_expo}) increases with $t$ and convexly decreases with $k$.
\end{pro}

The proof of Proposition \ref{pro_pk_expo_go0} is given in Appendix \ref{app_pro_pk_expo}.

Proposition \ref{pro_pk_expo_go0} shows that the growth rate of the expected profit decreases with service capacity $k$ given the fixed hovering time $T$. This is because the partitioned hovering time $\frac{T}{k}$ for pricing an individual service capacity decreases with $k$ in average sense. Similarly, the growth rate of the expected profit decreases with hovering time $T$ given the fixed service capacity $k$. We can also see that Proposition \ref{pro_pk_expo_go0} is consistent with the numerical results under discrete-time case in Fig. \ref{fig_pricek}. Moreover, the optimal price increases faster as $k$ decreases due to the scarce service capacity to sell within $t$ time period. Thus, $p_k(t)$ convexly decreases with $k$. %increases with $t$ as the UAV has more leftover time to encountering more users while decreases with $k$ due to smaller service capacity. Moreover, the optimal price increases faster as $k$ decreases.the properties of the optimal price $p_k(t)$ under continuous-time case is consistent with those under discrete-time case shown in Fig. \ref{fig_pricek}.

%\begin{figure}
%\centering\includegraphics[scale=0.35]{benchmark1}\caption{Ratio of the expected profit under incomplete information and complete information $R_k(T)/\hat{R}_k(T)$ versus hovering time $T$ and service capacity $k$.}\label{fig_benchmark_compare_profit}
%\end{figure}

\begin{figure}
\centering\includegraphics[scale=0.33]{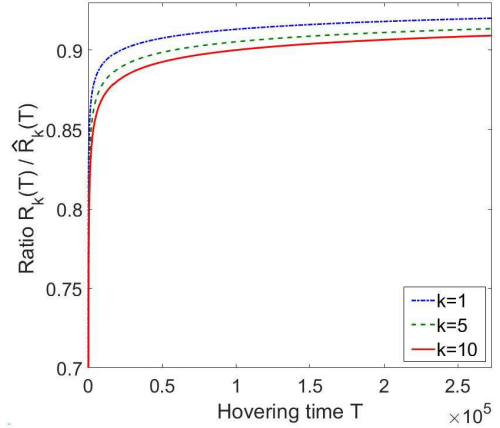}\caption{Ratio of the expected profits under incomplete information and complete information $R_k(T)/\hat{R}_k(T)$ versus hovering time $T$ and service capacity $k$.}\label{fig_benchmark_compare_profit}
\end{figure}

\subsection{Comparison with complete information benchmark}

We have finished analyzing the dynamic pricing under incomplete information above. We wonder the performance gap with ideally complete information,
where the UAV can perfectly observe a user's service valuation $v$ upon his immediate arrival.
Still, the UAV cannot observe future users' arrival pattern or valuations, otherwise, the analysis is trivial and the result is impractical. According to the threshold-based assignment policies developed for complete information in \cite{gershkov2014dynamic} under continuous-time case, we have the following proposition.

%Then, we have the following proposition.

\begin{pro}\label{pro_ratio_to_1} For $k\in\{1,2,3\}$, $\lim_{T\rightarrow\infty}\frac{R_k(T)}{\hat{R}_k(T)}=1$, where $R_k(T)$ and $\hat{R}_k(T)$ are the expected profits under incomplete and complete information for exponential distribution of users service valuations, respectively.
\end{pro}

Actually, for any finite $k<\infty$, we can iteratively obtain $\hat{R}_k(T)$ in non-closed-form according to \cite{gershkov2014dynamic}. Then we can also show the convergence of $\frac{R_k(T)}{\hat{R}_k(T)}$ to $1$ in a numerical way. As shown in Fig. \ref{fig_benchmark_compare_profit}, $R_k(t)$ approaches $\hat{R}_k(t)$ if the hovering time $T$ is sufficiently large. Moreover, $R_k(t)$ converges faster to $\hat{R}_k(t)$ as the service capacity $k$ decreases. This is because, as $k$ decreases, the partitioned hovering time $\lfloor\frac{T}{k}\rfloor$ for pricing an individual service capacity increases in average sense and is easier to become sufficient.
%This is because, under complete information, the UAV charges the user his service valuation $v$.
%However, under incomplete information, there exists profit loss due to the gap between the price and service valuation, which increases as the service capacity increases.

%\begin{figure}
%\centering\includegraphics[scale=0.32]{diff_variance_uniform_compare_twoT}\caption{Expected profits under complete information and incomplete information versus the variance with fixed mean $10$ under uniform distribution.}\label{fig_fixmean_uniform}
%\end{figure}

\begin{figure}
\centering\includegraphics[scale=0.3]{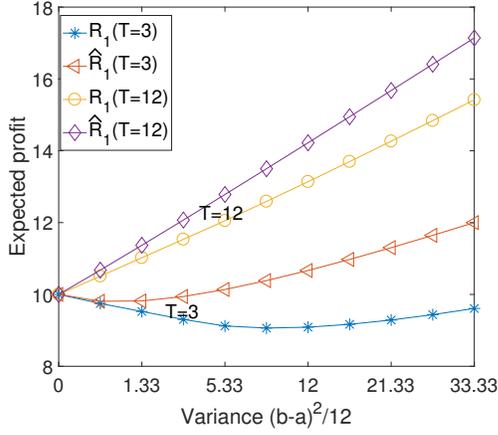}\caption{Expected profits under complete information and incomplete information versus the variance with fixed mean $10$ under uniform distribution.}\label{fig_fixmean_uniform}
\end{figure}

We also wonder how the expected profit changes with the user variance given fixed mean for the distribution of each user's service valuation. For the uniform distribution's CDF $F(v)=\frac{v-a}{b-a}, v\in[a,b]$, Fig. \ref{fig_fixmean_uniform} shows the maximum profits under both complete and incomplete information, and the former is greater. When the hovering time $T$ is big (e.g., $T=12$), the expected profits under incomplete and complete information always increase with the variance. This is because the users' maximum service valuation $b$ increases due to the increased variance and the UAV has enough time to wait for the user with higher service valuation to pay. However, when the hovering time $T$ is small (e.g., $T=3$), the expected profits first decrease with the variance due to the more information loss and then increase due to the larger upper bound $b$ to exploit. Note that even for the complete information, the UAV still has some information loss as it can only observe clearly an immediate arrival's valuation rather than any future information.

\section{UAV's Energy Allocation in Hovering Time and Service Capacity in Stage II}\label{sec_energyallocation}
%Resource Allocation at Stage I

Based on the analysis of optimal pricing in Section \ref{sec_pricing}, a longer hovering time $T$ results in a higher service price at the cost of smaller service capacity $k$. Therefore, in Stage $II$ the UAV under the total energy budget $B$ should balance $T$ and $k$ optimally for profit maximization. Its optimal energy allocation problem at the given hotspot is
\bee\label{equ_maxRkT} \max_{k, T\in\bZ^+}R_k(T), \ene
s.t. \bee\label{equ_maxcondition} T+ck\leq B, \ene
where $R_k(T)$ is returned by Algorithm \ref{alg1} for discrete-time case, and $B$ can be viewed as the maximum hovering time if the UAV does not use any energy to serve any user.
%the power consumption of serving one user is rescaled

At the optimality, (\ref{equ_maxcondition}) is tight to use up all the budget and the problem can be rewritten as
\bee\label{equ_R_k_allocation} \max_{k\in\bZ^+} R_k(B-ck). \ene

Recall that $\forall t<k, R_k(t)=R_{t}(t)$ in Proposition \ref{pro_Rktrelationship} and the UAV would not set $k$ to be larger than the maximum time, i.e., $k\leq B-ck$. Thus, integer decision $k$ is upper bounded by $\lfloor\frac{B}{1+c}\rfloor$. By using Algorithm \ref{alg1} for any $k\in\{1,...,\lfloor\frac{B}{1+c}\rfloor\}$, we can recursively calculate the corresponding expected profit $R_k(B-ck)$. Then, the UAV compares and chooses the best $k^*$ with maximal expected profit, i.e., $k^*=\arg\max_kR_k(B-ck)$.

\begin{figure}
\centering\includegraphics[scale=0.31]{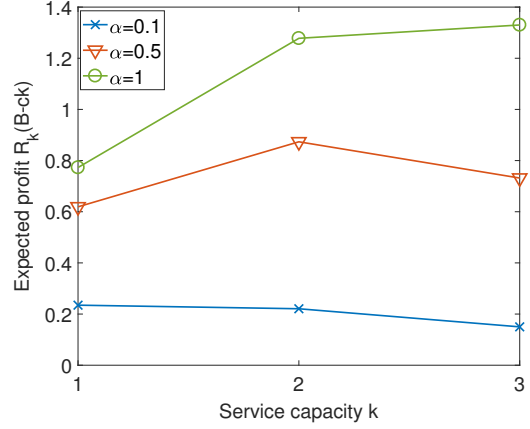}\caption{Expected profit versus service capacity $k$ for uniform service valuation distributions when $B=15$ and $c=3$. Here, integer decision $k$ is upper bounded by $\lfloor\frac{B}{1+c}\rfloor=3$.}\label{profit_k}
\end{figure}

Fig. \ref{profit_k} shows a numerical example for uniform distribution of users' service valuations under the discrete-time case.
\begin{itemize}
 \item For low user occurrence rate $\alpha$, it is better to only serve one user, i.e., $k^*=1$. It is worthwhile for the UAV to hover the longest possible time to encounter a user.
 \item For medium user occurrence rate $\alpha$, it is easier for the UAV to encounter more users and it should choose $k^*\in\{2,\cdots,\lfloor\frac{B}{c+1}\rfloor-1\}$, telling an optimal balance between encountered demand and service capacity supply.
 \item For high user occurrence rate $\alpha$, the UAV will meet many users and it will choose to serve as many users as possible, i.e., $k^*=\lfloor\frac{B}{c+1}\rfloor$.
\end{itemize}

To analytically obtain the energy allocation policy, similar to Section \ref{sec_continuous_pricing}, we apply continuous-time relaxation, where the maximum service capacity is $\lfloor \frac{B}{c}\rfloor$. Recall Proposition \ref{pro_pk_expo_go0} shows that $R_k(T)$ concavely increases with both $k$ and $T$ and $\alpha'$ is the arrival rate of Poisson process under the continuous-time model. Based on this result, we successfully develop an optimal energy allocation policy by assuming that the users' service valuations follow exponential distributions.%, as we can divide the time $B-ck_m$ into many time slots

%\begin{thm}\label{thm_tradeoff} The optimal service capacity $k^*$ depends on the user occurrence rate and is given as follows.
%\begin{itemize}
%  \item Low user occurrence regime ($\alpha'\leq\frac{2ce}{(B-2c)^2}$): the UAV will decide $k^*=1$ for serving one user only.
%  \item High user occurrence regime ($\alpha'>\frac{2ce}{(B-2c)^2}$): the UAV will decide \bee\label{equ_k^*} k^*=\arg\max_{k\in\bZ^+}\sum_{i=0}^k\frac{1}{i!}(\frac{\alpha' (B-ck)}{e})^i\in\{2,\cdots,\lfloor \frac{B}{c}\rfloor\}. \ene
%\end{itemize}
%\end{thm}

\begin{thm}\label{thm_tradeoff} The optimal service capacity $k^*$ depends on the user occurrence rate and is given as follows.
\begin{itemize}
  \item In the low user occurrence regime ($\alpha'\leq\frac{2ce}{(B-2c)^2}$), the UAV will decide $k^*=1$ for serving one user only.
  \item For medium user occurrence regime ($\frac{2ce}{(B-2c)^2}<\alpha'<\bar{\alpha}'$), the UAV will decide \bee\label{equ_k^*} k^*=\arg\max_{k\in\bZ^+}\sum_{i=0}^k\frac{1}{i!}(\frac{\alpha' (B-ck)}{e})^i\in\{2,\cdots,\lfloor \frac{B}{c}\rfloor-1\}. \ene
  \item For high user occurrence regime ($\alpha'\geq \bar{\alpha}'$), the UAV will decide $k^*=\lfloor \frac{B}{c}\rfloor$ for serving as many users as possible, where $\bar{\alpha}'$ is the unique solution to
      \bee\label{equ_largest_k_prove}\begin{split} &\frac{\alpha'}{e\lfloor \frac{B}{c}\rfloor !}(B-c\lfloor \frac{B}{c}\rfloor)^{\lfloor \frac{B}{c}\rfloor}-\sum_{i=1}^{\lfloor \frac{B}{c}\rfloor-1}(\frac{e}{\alpha'})^{\lfloor \frac{B}{c}\rfloor-i-1}\frac{1}{i!}\\
      &\times\Big((B-c(\lfloor \frac{B}{c}\rfloor-1))^i-(B-c\lfloor \frac{B}{c}\rfloor)^i\Big)=0. \end{split}\ene
\end{itemize}
The optimal hovering time is $T^*=B-ck^*$.
\end{thm}

The proof of Theorem \ref{thm_tradeoff} is given in Appendix \ref{app_thm_tradeoff}.
%The proof of Theorem \ref{thm_tradeoff} is given in Appendix \ref{app_thm_tradeoff}.

Note that the high user occurrence regime may not always exist since $\bar{\alpha}'$ as the solution to (\ref{equ_largest_k_prove}) can be infinity. It happens when $\frac{B}{c}$ is an integer and the hovering time $B-ck^*=B-c\frac{B}{c}$ is zero. In this case, we only have low and medium user occurrence regimes with $k^*<\frac{B}{c}$.

%For example, if $k=\lfloor \frac{B}{c}\rfloor$, there isn't enough hovering time for the UAV to encounter a user when $B-c\lfloor \frac{B}{c}\rfloor$ approaches $0$. Thus, the UAV will not choose the maximum possible service capacity.

The policy shown in Theorem \ref{thm_tradeoff} is threshold-based and easy to implement. As the total energy budget $B$ increases, the low user occurrence regime is less likely to happen and the optimal service capacity $k^*$ in (\ref{equ_k^*}) increases with $B$ and user occurrence rate $\alpha'$. This is because the UAV has more energy budget and thus it is better for it to serve more users. Moreover, we can see that Theorem \ref{thm_tradeoff}'s result under the continuous-time model is consistent with Fig. \ref{profit_k} under the discrete-time model. %Moreover, according to Fig. \ref{profit_k}, we can see that the optimal service capacity $k^*$ increases with user occurrence rate $\alpha$, preparing more supply to meet the increased demand.

% which is consistent with Fig. \ref{profit_k}.

\section{Optimal UAV Deployment in Stage I}\label{sec_UAVdeployment}

Given $M$ hotspots with heterogeneous user occurrence rates $\alpha_m$'s and distances $D_m$'s with $m\in\{1,...,M\}$ from the UAV station in Fig. \ref{fig_modelUAVs}, we now study how to deploy $N$ UAVs to such hotspots. Here each UAV has an identical initial energy budget $B_0$ fully charged at the UAV station. In the following, we will analyze the optimal UAV deployment strategy for a single UAV first and then extend to multiple cooperative UAVs.

\subsection{Deployment of a Single UAV to Heterogeneous Hotspots}

Given a single UAV's route going through $M'\leq M$ hotspots in sequence $\cH=\{H_1, H_2, \cdots, H_{M'}\}$, the route distance is $D_{H_1}+\sum_{m=1}^{M'-1}D_{H_m, H_{m+1}}$, where $D_{H_m, H_{m+1}}$ is the flying distance from hotspot $H_m$ to $H_{m+1}$. Given the initial energy budget $B_0$, the UAV spends energy traveling in the route and its remaining energy profile is partitioned among $M'$ hotspots in the meantime for serving users there, denoted as $\cB_{M'}=\{B_{H_1},...,B_{H_{M'}}\}$ with $\sum_{m=1}^{M'}B_{H_m}=B_0-D_{H_1}-\sum_{m=1}^{M'-1}D_{H_m, H_{m+1}}$. Here we normalize the energy consumption per unit flying distance as one. Given energy budget $B_{H_m}$ for hotspot $H_m$, we still need to decide the energy allocation to hovering time $T_{H_m}$ and service capacity $k_{H_m}$ in Stage II as well as the dynamic pricing during hovering time $T_{H_m}$ in Stage III. Given routing strategy $\cH$ and energy partition $\cB_{M'}$, based on (\ref{equ_R_k_allocation}) and Algorithm \ref{alg1}, we can obtain the UAV's overall expected profit by covering $M'$ hotspots in the whole route as
\bee\label{equ_singleUAVprofit} \Psi_{\cH}(\cH,\cB_{M'})=\sum_{H_m\in\cH}\max_{k_{H_m}\in\bZ^+}R_{k_{H_m}}^{\alpha_{H_m}}(B_{H_m}-ck_{H_m}), \ene
where $R_{k_{H_m}}^{\alpha_{H_m}}(t)$ is the expected profit from hotspot $H_m$ with user occurrence rate $\alpha_{H_m}$ for serving $k_{H_m}$ users at any time $t$.%\footnote{Here we normalize the energy consumption per flying distance as a unit.}

Still, the UAV needs to decide the route by considering $\sum_{M'=1}^MC_M^{M'} M'!$ possibilities, where the UAV needs to first choose $M'$ out of $M$ hotspots and each $M'$ introduces $M'!$ possible ordering sequences. Among these routing possibilities, the UAV finds the optimal route with hotspot sequence $\cH$ and energy allocation $\cB_{M'}$ to maximize the overall expected profit in (\ref{equ_singleUAVprofit}). %\bee \max_{\cH}\max_{\cB_{M'}}\Psi_{\cH}(\cB_{M'}). \ene

This routing problem followed by energy allocation and pricing is complicated and we can only solve in a numerical way. For analysis tractability, we still apply continuous-time relaxation on the discrete-time horizon and analyze the optimal UAV routing by assuming exponential distributions for users' service valuations. We also consider the situation when the hovering time increases in the energy allocated to the hotspot and the optimal service capacity increases in the user occurrence rate for any given energy budget.\footnote{Based on our exhaustive simulation, this assumption will not change the result in Theorem \ref{thm_trajectory_one}.}

\begin{thm}\label{thm_trajectory_one} For any number $M$ of heterogeneous hotspots distributed on the ground plane, it is optimal to deploy the single UAV to only one hotspot denoted by $m^*=\arg\max_{m\in\{1,...,M\}}\max_{k_m} R_{k_m}^{\alpha_m}(B_0-D_m-ck_m)$ without serving any other hotspot in the route.
\end{thm}

The proof of Theorem \ref{thm_trajectory_one} is given in Appendix \ref{app_thm_singleUAV}, and we use the results of Theorem \ref{thm_tradeoff} in Stage $II$ and (\ref{equ_Rk_expo_go0}) in Stage $III$ for the proof. %given in Appendix \ref{app_thm_singleUAV}.

Theorem \ref{thm_trajectory_one} tells that the single UAV will only serve the first best hotspot with maximum individual expected profit. If a part of energy budget is removed from serving the first best hotspot to also serve the second best, the overall profit decreases as the UAV's marginal profit from serving the second best hotspot is lower. Even if the UAV has to bypass a hotspot before reaching the first best hotspot, it will not spend any hovering time or service capacity there in the mean time.

Without loss of generality, we sort the $M$ heterogeneous hotspots according to their individual expected profits served by the single UAV, i.e., \bee\label{equ_sort_hotspots} R_{k_1^*}^{\alpha_1}(B_0-D_1-ck_1^*)\geq\cdots\geq R_{k_M^*}^{\alpha_M}(B_0-D_M-ck_M^*), \ene
where
\bee\label{equ_21} k_m^*=\arg\max_{k_m\in\bZ^+}R_{k_m}^{\alpha_m}(B_0-D_m-ck_m), \ene
$m\in\{1,\cdots,M\}$, and now hotspot $1$ is the first best followed by hotspot $2$ and others. Other than profit maximization objective, if the UAV further has the fairness commitment constraint to serve a minimum number of hotspots with certain hovering time in each, then we can still apply the sorting in (\ref{equ_sort_hotspots}) for choosing profitable hotspots and deciding the energy allocation differently among them.

%This is because the marginal profit reduced from the best hotspots is larger than that gained from the second-best hotspots if certain energy budget is shifted from the best hotspot to the second-best hotspot. Thus, it is not worthwhile to deploy the UAV to more than one hotspot.

\begin{algorithm}[t]
\caption{Deployment of $N$ UAVs to $M$ hotspots}
\begin{algorithmic}[1]

%\STATE \textbf{Input:}
%Number of UAVs $N$, number of hotspots $M$, all efficient UAV deployment set $\cA(\cN)$, each UAV's total energy budget $B_0$, energy consumption for serving a user $c$, distribution of users' service valuation $F(v)$, hotspot $m$'s user occurrence rate $\alpha_m$ and flying distance $D_m$ from the UAV station, $m\in\{1,...,M\}$
%\STATE \textbf{Output:}
%Optimal UAV deployment $\cN^*$

\FOR{Any UAV deployment profile $\cN$}
\FOR {$m=1$ to $M$}
\IF {$n_m=0$}
\STATE $R_{k_{m}}^{\alpha_m}=0$
\ELSE
\FOR {$k_{m} = 1$ to $\lfloor\frac{B_0-D_m}{1+\frac{c}{n_m}}\rfloor$ }
%\STATE $R_{k_{i}}^{\alpha_i}(0)=0$
\FOR {$t = 1$ to $\lfloor B_0-D_m-\frac{ck_m}{n_m}\rfloor$ }
\IF {$t\leq k_m-1$}
	\STATE {$R_{k_{m}}^{\alpha_m}(t)=R_{t}^{\alpha_m}(t)$}
\ELSE
    \STATE {compute $p_{k_m}^{\alpha_m}(t)$ as the unique solution to (\ref{equ_p_k(t)});}
	\STATE {update $R_{k_{m}}^{\alpha_m}(t)$ according to $p_{k_m}^{\alpha_m}(t)$ and (\ref{equ_R_k(t)});}
\ENDIF
\ENDFOR
\RETURN $R_{k_{m}}^{\alpha_m}(\lfloor B_0-D_m-\frac{ck_m}{n_m}\rfloor)$\\
\ENDFOR
\RETURN $\max_{k_{m}}R_{k_{m}}^{\alpha_m}(\lfloor B_0-D_m-\frac{ck_m}{n_m}\rfloor)$\\
\ENDIF
\ENDFOR
\RETURN $\Phi(\cN)$
\ENDFOR
\RETURN $\cN^*=\arg\max_{\cN}\Phi(\cN)$

\end{algorithmic}
\label{alg3}
\end{algorithm}

\subsection{Deployment of Multiple UAVs to Hotspots: Forking or Not}

\newcounter{myeq1}
\begin{figure*}[ht]
\setcounter{myeq1}{\value{equation}}
\setcounter{equation}{22}
\bee\label{equ_thm_forking} \varphi=\frac{\max_{k_1}\sum_{i=0}^{k_{1}}\frac{1}{i!}(\frac{\alpha'_1 (B_0-D_1-\frac{ck_{1}}{N})}{e})^i-\max_{k_1}\sum_{i=0}^{k_{1}}\frac{1}{i!}(\frac{\alpha'_1 (B_0-D_1-\frac{ck_{1}}{N-1})}{e})^i}{(\max_{k_1}\sum_{i=0}^{k_{1}}\frac{1}{i!}(\frac{\alpha'_1 (B_0-D_1-\frac{ck_{1}}{N-1})}{e})^i)(\sum_{i=1}^{k_{2}^*}\frac{1}{i!}(\frac{\alpha'_1 (B_0-D_2-ck_2^*)}{e})^i)}.\ene
\setcounter{equation}{\value{myeq1}}
\hrulefill
\end{figure*}

We are now ready to study how to assign multiple UAVs simultaneously from the common UAV station to heterogeneous hotspots, and we want to answer the key question: whether should all the UAVs still be deployed to the first best hotspot only as in Theorem \ref{thm_trajectory_one} or should they fork to serve different hotspots.\footnote{We assume UAVs are deployed at the same time to rapidly provide users with UPS. Otherwise, for profit maximization purpose, all the UAVs will be deployed one by one without any overlap to independently serve the first best hotspot only, as in Theorem \ref{thm_trajectory_one}.} It is possible for more than one UAV to cooperatively serve the same hotspot by ``pooling" their service capacities, e.g., when only one hotspot is close and reachable within the energy budget $B_0$. Given a number $n_m$ of UAVs cooperatively serving hotspot $m, m\in\{1,\cdots,M\}$, they will stay for the same amount of hovering time $T_m$ due to their symmetry. To keep the same energy consumption rates during hovering, $n_m$ UAVs take turns to serve users. For example, in the caching application, each UAV can sequentially distribute $1/n_m$ segment of the requested popular file to the same user. According to Theorem \ref{thm_trajectory_one}, the same group of $n_m$ UAVs will only travel to one hotspot. It is possible that $n_m=0$ or $1$. For a particular hotspot $m$, the objective of $n_m$ cooperative UAVs deployed there is to maximize their total expected profit,
\bee\label{equ_Rk_ni} \max_{k_{m},T_m\in\bZ^+}R_{k_{m}}^{\alpha_m}(T_m), \ene
s.t. \bee\label{equ_multiUAVcondition} n_mT_m+ck_{m}\leq n_m(B_0-D_m), \ene
where they decide total service capacity $k_m$ jointly in Stage $II$ by pooling their residual energy $n_m(B_0-D_m)$ upon reaching hotspot $m$. The cooperation among UAVs helps pool their service capacities to jointly decide pricing in Stage $III$, yet they waste more energy in hovering at the same time. %This is different from our analysis for $N=1$ UAV in Section \ref{sec_continuous_pricing}.

At the optimality, (\ref{equ_multiUAVcondition}) should be tight. Then, the overall expected profit given the UAV deployment profile $\cN=\{n_1,...,n_M\}$ to $M$ hotspots with $\sum_{m=1}^M n_m=N$ is
\bee\label{equ_PhiN} \Phi(\cN)=\sum_{m=1}^M\max_{k_{m}\in\bZ^+}R_{k_{m}}^{\alpha_m}( B_0-D_m-\frac{ck_m}{n_m}). \ene
Note that $R_{k_{m}}^{\alpha_m}=0$ for hotspot $m$ if $n_m=0$. For the discrete time model, the service capacity $k_m$ at hotspot $m$ should be no larger than the maximum hovering time, i.e., $k_m\leq B_0-D_m-\frac{ck_{m}}{n_m}$. Thus, $k_m\leq\lfloor\frac{B_0-D_m}{1+\frac{c}{n_m}}\rfloor$.

%Recall that we have sorted the $M$ heterogeneous hotspots according to (\ref{equ_sort_hotspots}). Then, given $N\geq 2$ UAVs for $M\geq 2$ hotspots, we only need to consider the UAV deployment profile $\cN=\{n_1,...,n_M\}$ that satisfies $n_1\geq n_2\geq\cdots\geq n_M$ and $\sum_{m=1}^M n_m=N$, and we denote this efficient subset of $\cN$ as $\cA(\cN)$.

In Algorithm \ref{alg3} given any UAV deployment profile $\cN$, we first compute the expected profit $\max_{k_{m}}R_{k_{m}}^{\alpha_m}(\lfloor B_0-D_m-\frac{ck_m}{n_m}\rfloor)$ under optimal energy allocation for each hotspot $m$. Then, by comparing the overall expected profits $\Phi(\cN)$ in (\ref{equ_PhiN}) under all possible efficient UAV deployment profiles $\cA(\cN)$, the optimal UAV deployment $\cN^*=\arg\max_{\cN\in\cA(\cN)}\Phi(\cN)$ can be obtained. The computation complexity of Algorithm \ref{alg3} is $O(N^MM\max_{m\in\{1,...,M\}}(B_0-D_m)^2)$.

\begin{figure*}[t]
\centering
\subfigure[Optimal UAV deployment profile $\cN^*=\{3,2,0,0,0\}$ among the five hotspots.]{\label{subfig_2}
\begin{minipage}{.31\textwidth}
\includegraphics[width=1\textwidth]{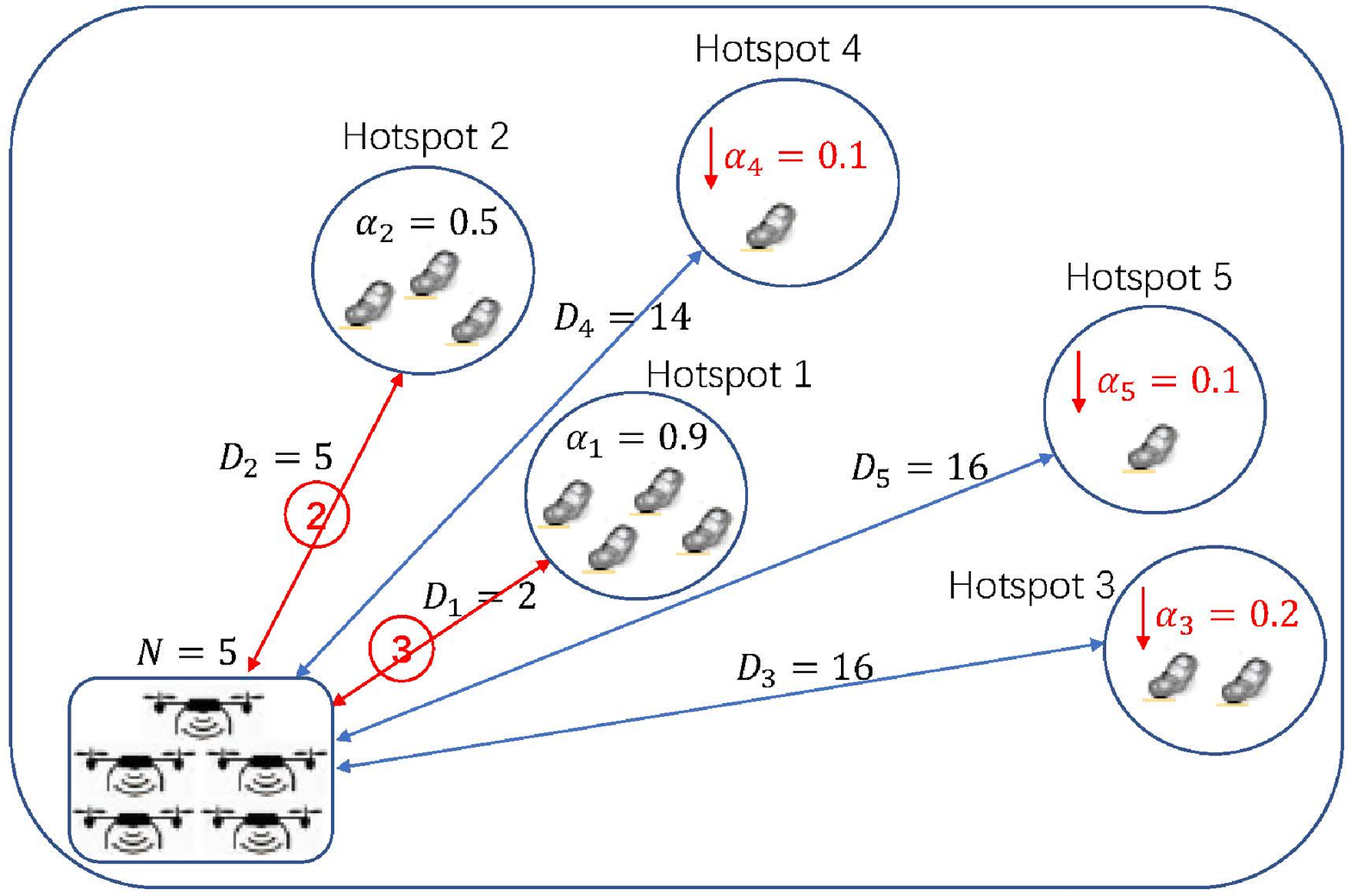}
\end{minipage}
}
\subfigure[Optimal UAV deployment profile $\cN^*=\{5,0,0,0,0\}$ among the five hotspots.]{\label{subfig_3}
\begin{minipage}{.31\textwidth}
\includegraphics[width=1\textwidth]{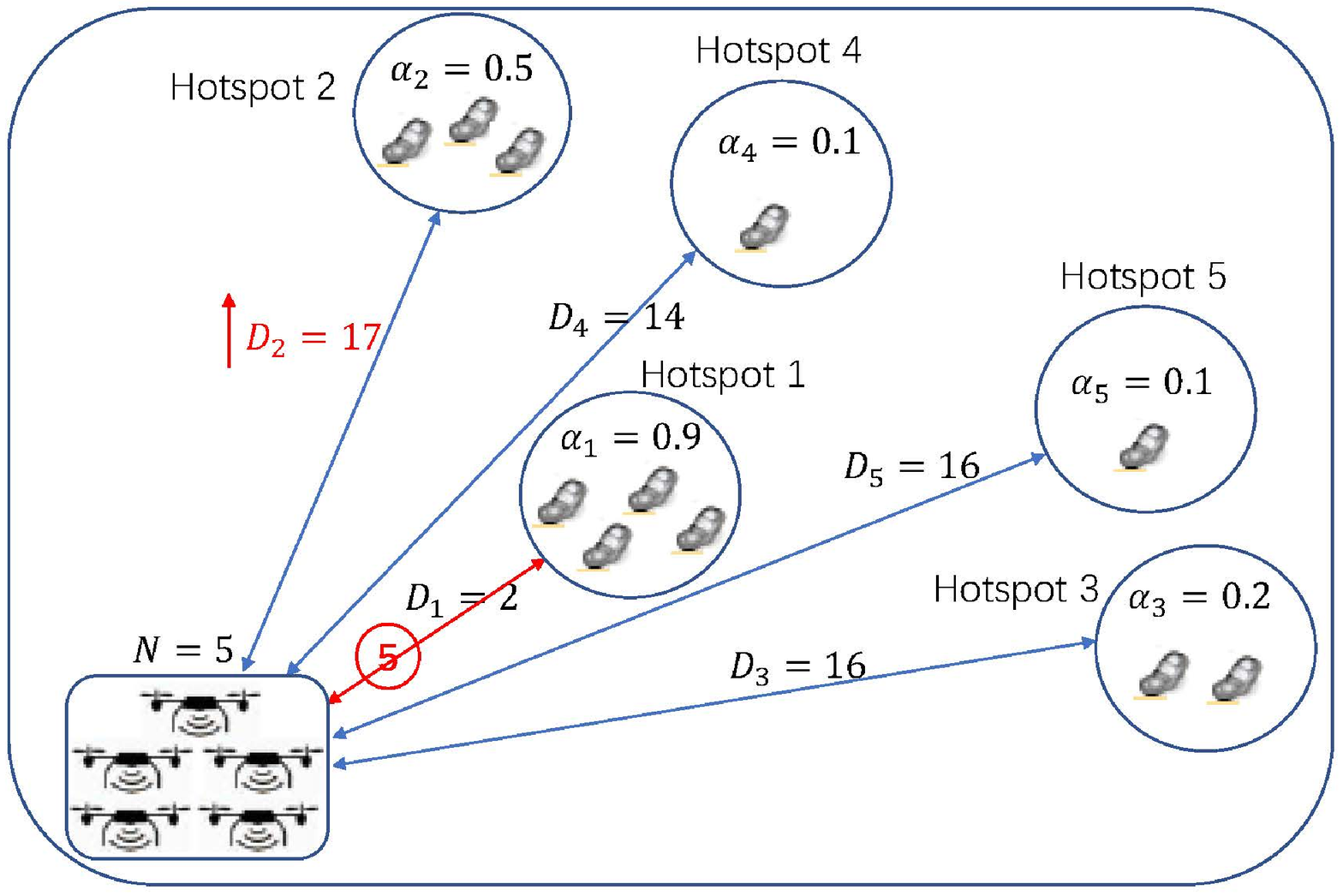}
\end{minipage}
}
\subfigure[Optimal UAV deployment profile $\cN^*=\{5,2,1,1,0\}$ among the five hotspots.]{\label{subfig_4}
\begin{minipage}{.31\textwidth}
\includegraphics[width=1\textwidth]{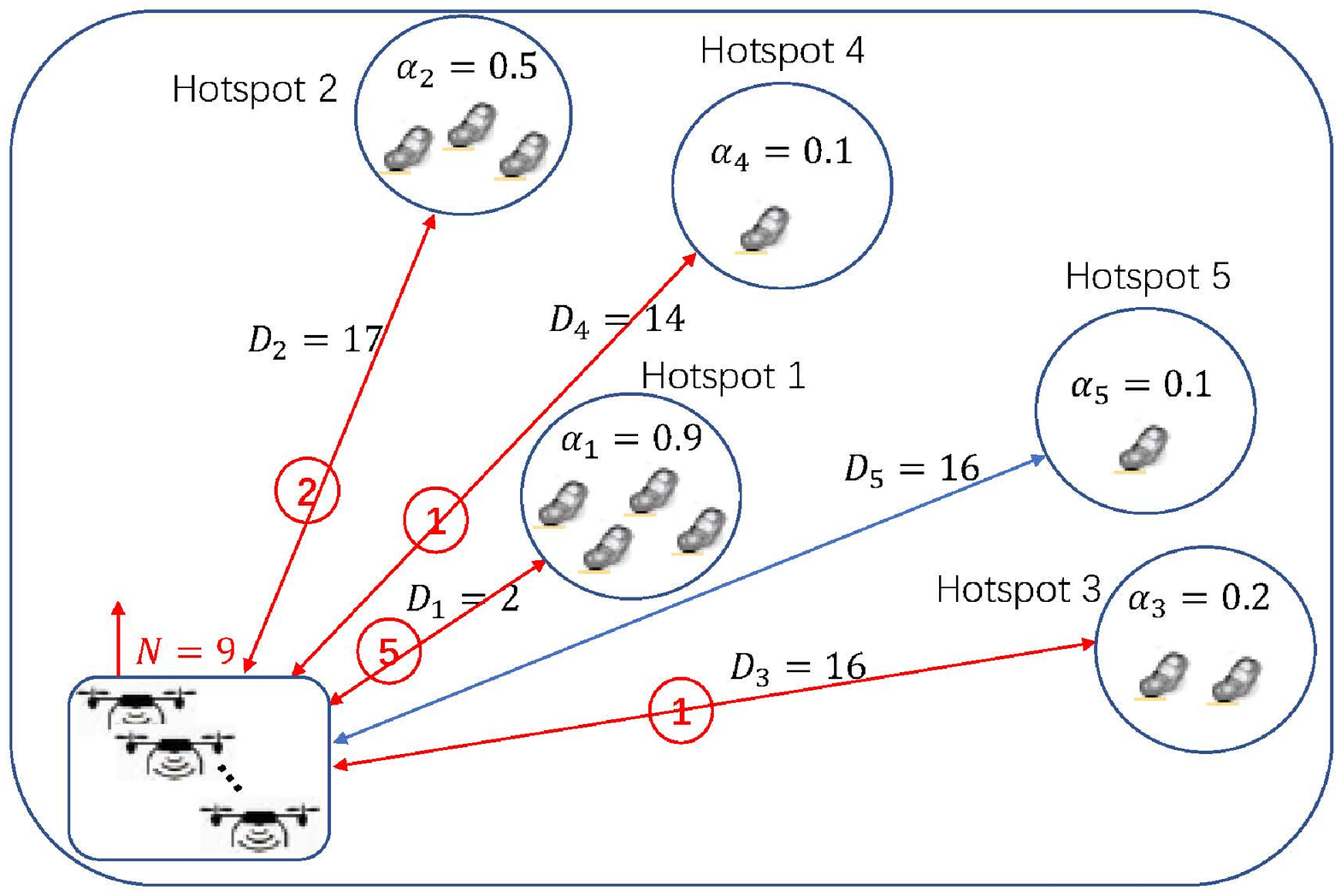}
\end{minipage}
}
\caption{Illustration of the optimal deployment of multiple UAVs to different hotspots when $B_0=20, c=2$. The number in red circle indicates how many UAVs are assigned to the corresponding hotspot.}\label{fig_exaUAVs}
\end{figure*}

%Under this order, hotspot $1$ is the first best and hotspot $2$ is the second best.

In the following, we further analyze whether UAVs should all center at the first best hotspot $1$ or fork to hotspot $2$ (or more hotspots), by assuming exponential distributions of users' service valuations for the relaxed continuous-time model with $\alpha'$ as the user arrival rate of Poisson process.

\begin{pro}\label{pro_forking} Given $N\geq 2$ UAVs for $M\geq 2$ hotspots, the UAVs will fork in their deployment to serve different hotspots rather than all centering at the first best hotspot $1$ if \bee\label{equ_forking1} \frac{\alpha'_2}{\alpha'_1}>\max(\varphi^{\frac{1}{k_2^*}},\varphi),\ene
where $\varphi$ is given in (\ref{equ_thm_forking}) and $k_2^*$ is given in (\ref{equ_21}). The forking condition in (\ref{equ_forking1}) is more likely to hold for a smaller flying distance $D_2$ to hotspot $2$ or a larger user occurrence rate $\alpha'_2$.
\end{pro}
%{\tiny\bee\label{equ_thm_forking}\begin{split} \varphi=&\frac{\max_{k_1}\sum_{i=0}^{k_{1}}\frac{1}{i!}(\frac{\alpha'_1 (B_0-D_1-\frac{ck_{1}}{N})}{e})^i}{(\max_{k_1}\sum_{i=0}^{k_{1}}\frac{1}{i!}(\frac{\alpha'_1 (B_0-D_1-\frac{ck_{1}}{N-1})}{e})^i)(\sum_{i=1}^{k_{2}^*}\frac{1}{i!}(\frac{\alpha'_1 (B_0-D_2-ck_2^*)}{e})^i)}\\
%&-\frac{1}{\sum_{i=1}^{k_{2}^*}\frac{1}{i!}(\frac{\alpha'_1 (B_0-D_2-ck_2^*)}{e})^i}.\end{split}\ene}
%in (\ref{equ_thm_forking})% i.e., the user occurrence rate $\alpha_2'$ of the second best hotspot becomes larger or the distance $D_2$ of the second best hotspot becomes shorter.

\newcounter{mytempeqncnt1}
\setcounter{mytempeqncnt1}{\value{equation}}
\setcounter{equation}{23}

The proof of Proposition \ref{pro_forking} is given in Appendix \ref{app_pro_forking}.

According to Proposition \ref{pro_forking}, we can see that the UAVs are more likely to fork to serve hotspot $2$ (and others) if the latter has a similar user density $\alpha'_2$ and flying distance $D_2$ as hotspot $1$.

We also wonder the impact of the number of UAVs on the forking deployment. As a numerical example, we first apply Algorithm \ref{alg3} to consider $N=5$ UAVs to be deployed to $M=5$ hotspots in Fig. \ref{fig_modelUAVs}, Fig. \ref{subfig_2}, and Fig. \ref{subfig_3} under the same setup. Later in Fig. \ref{subfig_4}, we will increase the number of UAVs from $5$ to $9$. We consider users' valuations follow the exponential distribution with $\lambda=1$.
\begin{itemize}
  \item In Fig. \ref{fig_modelUAVs}, it is optimal to deploy $5$ UAVs to all $5$ hotspots. As the user occurrence rates of hotspots $3,4,5$ decrease from Fig. \ref{fig_modelUAVs} to Fig. \ref{subfig_2}, we will not serve these hotspots but deploy $3$ UAVs to hotspot $1$ and $2$ UAVs to hotspot $2$. Now hotspot $1$ ($2$) is the first (second) best.
  \item As the flying distance to hotspot $2$ increases from Fig. \ref{subfig_2}'s $D_2=5$ to Fig. \ref{subfig_3}'s $D_2=17$, all the UAVs will stop forking and only serve hotspot $1$ without considering hotspot $2$. This is consistent with Proposition \ref{pro_forking}.
  \item Finally, Fig. \ref{subfig_4} shows that as the number of UAVs increases from Fig. \ref{subfig_3}, the UAVs fork to serve different hotspots again. This is because when many cooperative UAVs center at the same hotspot, they waste a lot of energy in hovering for the same group of users as shown in (\ref{equ_multiUAVcondition}). Therefore, it is better for some UAVs to fork to serve different hotspots (though distant) and meet more demands.
\end{itemize}

\section{Conclusion}\label{sec_conclusion}

%In this paper, we first analyze the UAV's dynamic pricing under incomplete information including random user arrivals and unknown service valuations. It is shown that the UAV should ask for a higher price if the leftover hovering time is longer or its service capacity is smaller, and its expected profit approaches to that under complete user information if the hovering time is sufficiently large. Then, given a hotspot, the energy allocation to hovering time and service capacity is optimized. We show that, given the energy budget at the hotspot, a shorter hovering time or a larger service capacity should be allocated as the hotspot's mean user occurrence rate increases. For the UAV deployment, we prove that it is optimal for a single UAV to only serve the best hotspot. While for multiple UAVs, they prefer to fork to serve different hotspots when hotspots are more symmetric or the UAV number is large.

In this paper, we first analyze the UAV's dynamic pricing under incomplete information including random user arrivals and unknown service valuations. Then, given a hotspot to deploy, the energy allocation to hovering time and service capacity is optimized. Finally, we show the optimal UAVs' deployment to potential hotspots.

There are some possible directions to study in the future. For example, now we plan the UAVs' dynamic pricing and energy allocation strategies beforehand at the UAV station, which saves the implementation complexity for UAVs to decide and operate in real time. Yet given more artificial intelligence, the UAVs in the future may learn and adapt prices to realized user occurrence over time and decide to hover longer or not. Another future direction is to decide the UPS provision under minimum service requirements (e.g., minimum number of hotspots or demands to serve) for fairness concern besides profitability objective.

\appendices

\section{Proof of Proposition \ref{pro_priceincrease}}\label{app_pro_priceincrease}

Under the reasonable assumption of regularity, $R_k(t)$ in (\ref{equ_R_k(t)}) is a concave function with respect to $p_k(t)$, and the solution $p_k(t)$ to (\ref{equ_p_k(t)}) is the unique optimal price. Algorithm \ref{alg1} uses this result and the resulting dynamic pricing strategy is optimal. As shown in Algorithm \ref{alg1}, given the service capacity $k$, for each $j$th-to-last user ($j=1,...,k$), we should compute the optimal price $p_j(t)$ and expected profit $R_j(t)$ at each time slot $t\in\{1,...,T\}$ during the hovering time $T$. Therefore, the computation complexity of Algorithm \ref{alg1} is $O(kT)$.

Then, we prove that $p_1(t)$ is a non-decreasing function of $t$. If $k=1$, we can simplify (\ref{equ_p_k(t)}) and the optimal price $p_1(t)$ is the unique solution to \bee\label{equ_optimalp1} p_1(t)-\frac{1-F(p_1(t))}{f(p_1(t))}=R_1(t-1). \ene Since $\phi(p_1(t))$ as the left-hand-side of (\ref{equ_optimalp1}) increases with $p_1(t)$, the optimal price $p_1(t)$ increases with $R_1(t-1)$. Note that $R_1(t)\geq R_1(t-1)$. Therefore, $p_1(t)$ is a non-decreasing function of $t$.

Finally, we prove that $p_1(t)$ is a non-decreasing function of $\alpha$. According to (\ref{equ_optimalp1}), the optimal expected profit in (\ref{equ_R1(t)}) can be rewritten as
\bee\begin{split} R_1(t)=p_1(t)-\frac{1-F(p_1(t))}{f(p_1(t))}(1-\alpha(1-F(p_1(t)))).\end{split}\ene
Taking the derivative of $R_1(t)$ with respect to $p_1(t)$, we have
\bee\label{equ_prove7}\begin{split} \frac{dR_1(t)}{dp_1(t)}=&(1-\alpha(1-F(p_1(t))))\\
&\times(2+\frac{(1-F(p_1(t)))f'(p_1(t))}{f^2(p_1(t))}). \end{split}\ene
Since $\phi(p_1(t))$ increases with $p_1(t)$, i.e., $\frac{d\phi(p_1(t))}{dp_1(t)}>0$, we have $2+\frac{(1-F(p_1(t)))f'(p_1(t))}{f^2(p_1(t))}>0$ in (\ref{equ_prove7}). Note that $1-\alpha(1-F(p_1(t)))>0$. Therefore, $\frac{dR_1(t)}{dp_1(t)}>0$, i.e., $R_1(t)$ increases with $p_1(t)$.

Since $R_1(0)=0$, by optimizing $R_1(1)=\alpha p_1(1)(1-F(p_1(1)))$, we can see that the optimal price $p_1(1)$ is not a function of $\alpha$ and $R_1(1)$ linearly increases with $\alpha$. According to (\ref{equ_optimalp1}), $p_1(2)$ increases with $R_1(1)$ as $p_1(t)-\frac{1-F(p_1(t))}{f(p_1(t))}$ increases with $p_1(t)$, which means that $p_1(2)$ also increases with $\alpha$. As we discussed in the last paragraph, $R_1(2)$ increases with $p_1(2)$, which means that $R_1(2)$ also increases with $\alpha$. Similarly, according to (\ref{equ_optimalp1}), $p_1(3)$ increases with $R_1(2)$, which means that $p_1(3)$ increases with $\alpha$. An iterative analysis shows that $p_1(t)$ for any $t$ is a non-decreasing function of $\alpha$.

\section{Proof of Proposition \ref{pro_pk_expo_go0}}\label{app_pro_pk_expo}

Note that $\frac{\sum_{i=0}^k\frac{1}{i!}(\frac{\alpha' t}{e})^i}{\sum_{i=0}^{k-1}\frac{1}{i!}(\frac{\alpha' t}{e})^i}=1+\frac{\frac{1}{k!}(\frac{\alpha' t}{e})^k}{\sum_{i=0}^{k-1}\frac{1}{i!}(\frac{\alpha' t}{e})^i}=1+\frac{\frac{1}{k!}(\frac{\alpha'}{e})^k}{\sum_{i=0}^{k-1}\frac{1}{i!}(\frac{\alpha'}{e})^i\frac{1}{t^{k-i}}}$, which is increasing in $t$. Then, according to (\ref{equ_pk_expo}), we have $p_k(t)$ increases with $t$.

In the following, we will prove that $p_k(t)$ decreases with $k$. To prove $p_k(t)\geq p_{k+1}(t)$, we only need to prove \bee \frac{\sum_{i=0}^k\frac{1}{i!}(\frac{\alpha' t}{e})^i}{\sum_{i=0}^{k-1}\frac{1}{i!}(\frac{\alpha' t}{e})^i}\geq \frac{\sum_{i=0}^{k+1}\frac{1}{i!}(\frac{\alpha' t}{e})^i}{\sum_{i=0}^{k}\frac{1}{i!}(\frac{\alpha' t}{e})^i}, \ene
which is equivalent to prove
\bee\begin{split}\label{equ_prove_p_increase_k} &\frac{1}{k!}(\frac{\alpha' t}{e})^k\frac{1}{(k+1)!}(\frac{\alpha' t}{e})^{k+1}\\
\geq &(\frac{1}{(k+1)!}(\frac{\alpha' t}{e})^{k+1}-\frac{1}{k!}(\frac{\alpha' t}{e})^k)\sum_{i=0}^k\frac{1}{i!}(\frac{\alpha' t}{e})^i. \end{split}\ene
Multiply both side of (\ref{equ_prove_p_increase_k}) with $(k+1)!$ and $(\frac{e}{\alpha' t})^k$, we can rewrite (\ref{equ_prove_p_increase_k}) as
\bee\begin{split}\label{equ_prove_p_increase_k2} (k+1)\sum_{i=0}^k\frac{1}{i!}(\frac{\alpha' t}{e})^i\geq &\sum_{i=0}^{k-1}\frac{1}{i!}(\frac{\alpha' t}{e})^{i+1}\\
=&\sum_{i=1}^k\frac{1}{(i-1)!}(\frac{\alpha' t}{e})^i. \end{split}\ene

Note that $(k+1)\sum_{i=0}^k\frac{1}{i!}(\frac{\alpha' t}{e})^i>\sum_{i=0}^k\frac{1}{(i-1)!}(\frac{\alpha' t}{e})^i>\sum_{i=1}^k\frac{1}{(i-1)!}(\frac{\alpha' t}{e})^i$. Therefore, (\ref{equ_prove_p_increase_k2}) always holds, which means the optimal price decreases with $k$. We can also verify that $2p_k(t)\leq p_{k+1}(t)+p_{k-1}(t)$, which shows that the optimal price decreases with $k$ convexly.

To prove that $R_k(t)$ is a concave with $k$, we only need to show that
\bee\label{equ_prove_Rk_concave_k}  R_{k+1}(t)+R_{k-1}(t)\leq 2R_k(t). \ene
Insert (\ref{equ_Rk_expo_go0}) into (\ref{equ_prove_Rk_concave_k}), it is equivalent to prove
\been\begin{split} &\frac{1}{k!}(\frac{\alpha' t}{e})^k\frac{1}{(k+1)!}(\frac{\alpha' t}{e})^{k+1}\\
\geq &(\frac{1}{(k+1)!}(\frac{\alpha' t}{e})^{k+1}-\frac{1}{k!}(\frac{\alpha' t}{e})^k)\sum_{i=0}^k\frac{1}{i!}(\frac{\alpha' t}{e})^i, \end{split}\enen
which is same as (\ref{equ_prove_p_increase_k}) and the proof is done.

Then, we will show that $R_k(t)$ is a concave with $t$, which is equivalent to show $\frac{\partial^2 R_k(t)}{\partial t^2}<0$. By taking the second derivative of $R_k(t)$ with respect to $t$, we have
\bee\begin{split} &\frac{\partial^2 R_k(t)}{\partial t^2}\\
=&\frac{\alpha'^2}{\lambda e^2}\frac{\sum_{i=0}^{k-2}\frac{1}{i!}(\frac{\alpha' t}{e})^i\sum_{i=0}^{k}\frac{1}{i!}(\frac{\alpha' t}{e})^i-(\sum_{i=0}^{k-1}\frac{1}{i!}(\frac{\alpha' t}{e})^i)^2}{(\sum_{i=0}^{k}\frac{1}{i!}(\frac{\alpha' t}{e})^i)^2}. \end{split}\ene

Since $R_k(t)$ is a concave with $k$, we have \bee \sum_{i=0}^{k-2}\frac{1}{i!}(\frac{\alpha' t}{e})^i\sum_{i=0}^{k}\frac{1}{i!}(\frac{\alpha' t}{e})^i-(\sum_{i=0}^{k-1}\frac{1}{i!}(\frac{\alpha' t}{e})^i)^2<0. \ene
Therefore, $\frac{\partial^2 R_k(t)}{\partial t^2}<0$ always holds.

\section{Proof of Theorem \ref{thm_tradeoff}}\label{app_thm_tradeoff}

As shown in Section \ref{sec_continuous_pricing}, the expected profit for hovering time $t$ and the corresponding optimal price are given in (\ref{equ_Rk_expo_go0}) and (\ref{equ_pk_expo}), respectively. When $\alpha'\leq\frac{2ce}{(B-2c)^2}$, we have $R_1(B-c)\geq R_2(B-2c)$. Then, we prove  $R_{k-1}(B-c(k-1))\geq R_k(B-ck)$ for any $k>2$ under the condition $\alpha'\leq\frac{2ce}{(B-2c)^2}$. Since \bee\begin{split}  &R_k(B-ck)-R_{k-1}(B-c(k-1))\\
=&\frac{1}{\lambda}\log(\frac{\sum_{i=0}^k\frac{1}{i!}(\frac{\alpha' (B-ck)}{e})^i}{\sum_{i=0}^{k-1}\frac{1}{i!}(\frac{\alpha' (B-c(k-1))}{e})^i}), \end{split}\ene
we only need to show \bee \sum_{i=1}^k\frac{1}{i!}(\frac{\alpha' (B-ck)}{e})^i\leq \sum_{i=1}^{k-1}\frac{1}{i!}(\frac{\alpha' (B-c(k-1))}{e})^i. \ene

For $k=3$, $B$ must be larger than $3c$. According to $\alpha'\leq\frac{2ce}{(B-2c)^2}$, we have \bee\begin{split} &\sum_{i=1}^3\frac{1}{i!}(\frac{\alpha' (B-3c)}{e})^i-\sum_{i=1}^{2}\frac{1}{i!}(\frac{\alpha' (B-2c)}{e})^i\\
\leq &\frac{2c^2(B-3c)^3}{3(B-2c)^4}-2c+\frac{c(B-3c)^2}{(B-2c)^2}\\
\leq &\frac{2c^2}{3(B-2c)}-c. \end{split}\ene

Since $B\geq 3c$, we have $\frac{2c^2}{3(B-2c)}-c<0$. Therefore, $R_2(B-2c)>R_3(B-3c)$.

For $k\geq 4$, i.e., $B$ must be larger than $4c$, we only need to prove if $\sum_{i=1}^{k-1}\frac{1}{i!}(\frac{\alpha' (B-c(k-1))}{e})^i\leq \sum_{i=1}^{k-2}\frac{1}{i!}(\frac{\alpha' (B-c(k-2))}{e})^i$, then $\sum_{i=1}^k\frac{1}{i!}(\frac{\alpha' (B-ck)}{e})^i\leq \sum_{i=1}^{k-1}\frac{1}{i!}(\frac{\alpha' (B-c(k-1))}{e})^i$.

Since $\sum_{i=1}^{k-1}\frac{1}{i!}(\frac{\alpha' (B-c(k-1))}{e})^i\leq \sum_{i=1}^{k-2}\frac{1}{i!}(\frac{\alpha' (B-c(k-2))}{e})^i$, we can show that

\bee\label{equ_1}\begin{split} &\sum_{i=1}^k\frac{1}{i!}(\frac{\alpha' (B-ck)}{e})^i- \sum_{i=1}^{k-1}\frac{1}{i!}(\frac{\alpha' (B-c(k-1))}{e})^i\\
\leq &\sum_{i=1}^{k-2}\frac{\alpha'^{i+1}}{ke^{i+1} i!}\Big((B-c(k-1))\big((B-c(k-2))^i\\
&-(B-c(k-1))^i\big)\\
&-\frac{k}{i+1}((B-c(k-1))^{i+1}-(B-ck)^{i+1})\Big)-\frac{\alpha' c}{e}\\
< &\sum_{i=1}^{k-2}\frac{\alpha'}{e k}\Big(\frac{(2c)^i}{(B-2c)^{2i} i!}(B-c(k-1))\\
&\times((B-c(k-2))^i-(B-c(k-1))^i)-c\Big)\\
< &\sum_{i=1}^{k-2}\frac{\alpha'}{e k}\Big(\frac{(2c)^i}{(B-2c)^{2i} i!}((B-2c)^{i+1}-(B-c(k-1))^{i+1})\\
&-c\Big)\\
<&\sum_{i=1}^{k-2}\frac{\alpha'}{e k}\Big(\frac{(2c)^i}{(B-2c)^{i-1} i!}-c\Big)
\end{split}\ene

Note that $B\geq 4c$. Thus, $2^ic^{i}-i!c(B-2c)^{i-1}\leq 2^ic^{i}-i!2^{i-1}c^i=2^ic^{i}(1-\frac{i!}{2})$. Since $1-\frac{i!}{2}\leq 0$ for $i\geq 2$, we have $\frac{(2c)^i}{(B-2c)^{i-1} i!}-c\leq0$ for any $i\geq 2$. When $i=1$, it is easy to check that the third equation of (\ref{equ_1}) is negative. Therefore, we have $R_{k-1}(B-c(k-1))\geq R_k(B-ck)$ for any $k>1$ when $\alpha'\leq\frac{2ce}{(B-2c)^2}$.

Then, we consider the case when $\alpha'>\frac{2ce}{(B-2c)^2}$, i.e., $R_1(B-c)<R_2(B-2c)$. Since the left-hand side of equation (\ref{equ_largest_k_prove}) is increasing with $\alpha'$ and the right-hand side is decreasing with $\alpha'$, the solution $\bar{\alpha}'$ is unique. Note that (\ref{equ_largest_k_prove}) can be rewrite as
\bee\begin{split} &\frac{1}{\lambda}\log(\sum_{i=0}^{\lfloor \frac{B}{c}\rfloor}\frac{1}{i!}(\frac{\alpha' (B-c\lfloor \frac{B}{c}\rfloor)}{e})^i)\\
=&\frac{1}{\lambda}\log(\sum_{i=0}^{\lfloor \frac{B}{c}\rfloor-1}\frac{1}{i!}(\frac{\alpha' (B-c(\lfloor \frac{B}{c}\rfloor-1))}{e})^i). \end{split}\ene
Therefore, when $\alpha'\geq \bar{\alpha'}$, we have $R_{\lfloor \frac{B}{c}\rfloor}(B-c\lfloor \frac{B}{c}\rfloor)\geq R_{\lfloor \frac{B}{c}\rfloor-1}(B-c(\lfloor \frac{B}{c}\rfloor-1))$. Then, we prove $R_{k-1}(B-c(k-1))\geq R_{k-2}(B-c(k-2))$ for any $k$ given $R_k(B-ck)\geq R_{k-1}(B-c(k-1))$.

Since $R_k(B-ck)\geq R_{k-1}(B-c(k-1))$ is equivalent to
\bee\begin{split} &\frac{1}{k!}(\frac{\alpha'}{e})^{k-1}(B-ck)^k\\
\geq&\sum_{i=1}^{k-1}\frac{1}{i!}(\frac{\alpha'}{e})^{i-1}\Big((B-c(k-1))^i-(B-ck)^i\Big), \end{split}\ene
we have
\bee\begin{split} &R_{k-1}(B-c(k-1))-R_{k-2}(B-c(k-2))\\
>&\sum_{i=1}^{k-2}(\frac{\alpha'}{e})^i\Big(\frac{k}{(i+1)!}((B-c(k-1))^i-(B-ck)^i)\\
&-\frac{1}{i!}((B-c(k-1))^i-(B-c(k-2))^i\Big)+\frac{ck}{B-ck}\\
>&\sum_{i=1}^{k-2}(\frac{\alpha'}{e})^i\frac{1}{i!}\Big((B-c(k-2))^i-(B-ck)^i\Big)+\frac{ck}{B-ck}\\
>&0.
\end{split}\ene
Therefore, we have $R_{k}(B-ck)\geq R_{k-1}(B-c(k-1))$ for any $k$ if $\alpha'\geq \bar{\alpha}'$.

If $\alpha'<\bar{\alpha}'$, we have $R_{\lfloor \frac{B}{c}\rfloor}(B-c\lfloor \frac{B}{c}\rfloor)< R_{\lfloor \frac{B}{c}\rfloor-1}(B-c(\lfloor \frac{B}{c}\rfloor-1))$. Since $R_1(B-c)<R_2(B-2c)$, there exists a $k^*\in\{2,\cdots,\lfloor \frac{B}{c}\rfloor-1\}$ such that $k^*=\arg\max_kR_k(B-ck)$.

%When $\alpha'>\frac{2ce}{(B-2c)^2}$, i.e., $R_1(B-c)<R_2(B-2c)$, it is manifest that $k^*\geq 2$.

\section{Proof of Theorem \ref{thm_trajectory_one}}\label{app_thm_singleUAV}

Given any two hotspots in the network, if the UAV will always serve only one hotspot rather than both of them, we can conclude that the UAV will always serve only one hotspot with maximum expected profit. Therefore, in the following, we will only consider two hotspots here and generally consider that the UAV will pass by closer hotspot $1$ first and then hotspot $2$. For any energy allocation to the hotspots $B_1$ and $B_2$, the optimal expected profit of hotspot $j, j\in\{1,2\}$ is
\bee  R_{\hat{k}_j}^{\alpha'_j}(B_j-c\hat{k}_j)=\frac{1}{\lambda}\log\Big(\sum_{i=0}^{\hat{k}_j}\frac{1}{i!}(\frac{\alpha'_j (B_j-c\hat{k}_j)}{e})^i\Big), \ene
where $\hat{k}_j=\arg\max_{k_j}R_{k_j}^{\alpha'_j}(B_j-ck_j),j\in\{1,2\}$.

First, we consider the case that $R_{k_1^*}^{\alpha'_1}(B_1^0-ck_1^*)\geq R_{k_2^*}^{\alpha'_2}(B_2^0-ck_2^*)$, i.e., the optimal expected profit of hotspot $1$ is larger than that of hotspot $2$ if the UAV chooses to serve only one hotspot, where $B_1^0=B_0-D_1, B_2^0=B_0-D_2$, and $k_i^*=\arg\max_{k_i}R_{k_i}^{\alpha'_i}(B_i^0-ck_i),i\in\{1,2\}$. Obviously, $D_{1,2}\geq B_1^0-B_2^0$.

Note that $R_{k_1^*}^{\alpha'_1}(B_1^0-ck_1^*)\geq R_{k_2^*}^{\alpha'_2}(B_2^0-ck_2^*)$ and $k_1^*, k_2^*$ are the optimal energy allocation, we have
\bee\begin{split}  \log(\sum_{i=0}^{k_1^*}\frac{1}{i!}(\frac{\alpha'_1 (B_1^0-ck_1^*)}{e})^i)
\geq&\log(\sum_{i=0}^{k_2^*}\frac{1}{i!}(\frac{\alpha'_2 (B_2^0-ck_2^*)}{e})^i)\\
\geq &\log(\sum_{i=0}^{k_1^*}\frac{1}{i!}(\frac{\alpha'_2 (B_2^0-ck_1^*)}{e})^i). \end{split}\ene

As the numbers of summation terms in $R_{k_1^*}^{\alpha'_i}(B_i^0-ck_1^*), i=1,2$ given in (\ref{equ_Rk_expo_go0}) are the same, we have $\frac{\alpha'_1 (B_1^0-ck_1^*)}{e}\geq\frac{\alpha'_2 (B_2^0-ck_1^*)}{e}$ for each summation term according to $R_{k_1^*}^{\alpha'_2}(B_2^0-ck_1^*)\leq R_{k_1^*}^{\alpha'_1}(B_1^0-ck_1^*)$. If $\alpha'_1<\alpha'_2$, we have $D_{1,2}\geq(B_2^0-ck_1^*)(\frac{\alpha'_2}{\alpha'_1}-1)$ due to $D_{1,2}\geq B_1^0-B_2^0$. Note that it is assumed that the hovering time increases in the energy allocated to the hotspot and the optimal service capacity $\bar{k}_i$ increases in $\alpha_i'$ for any given energy budget. Thus, $B_2^0-ck_1^*\geq B_2-c\bar{k}_1$ and $B_2-c\bar{k}_1\geq B_2-c\hat{k}_2$ due to $\alpha'_1<\alpha'_2$. Therefore, we have $D_{1,2}\geq(B_2^0-ck_1^*)(\frac{\alpha'_2}{\alpha'_1}-1)\geq (B_2-c\hat{k}_2)(\frac{\alpha'_2}{\alpha'_1}-1)$. By comparing each summation term in $R_{\hat{k}_2}^{\alpha'_1}(B_2+D_{1,2}-c\hat{k}_2)$ and $R_{\hat{k}_2}^{\alpha'_2}(B_2-c\hat{k}_2)$, we have
\bee\label{equ_proof_compare} R_{\hat{k}_2}^{\alpha'_1}(B_2+D_{1,2}-c\hat{k}_2)\geq R_{\hat{k}_2}^{\alpha'_2}(B_2-c\hat{k}_2). \ene

Since $\hat{k}_1, \hat{k}_2$ are the optimal energy allocation for each hotspot, we have
\bee\label{equ_compare_hat} R_{\hat{k}_1}^{\alpha'_1}(B_2+D_{1,2}-c\hat{k}_1)\geq R_{\hat{k}_2}^{\alpha'_1}(B_2+D_{1,2}-c\hat{k}_2). \ene
Thus, according to (\ref{equ_proof_compare}) and (\ref{equ_compare_hat}), we have $R_{\hat{k}_1}^{\alpha'_1}(B_2+D_{1,2}-c\hat{k}_1)\geq R_{\hat{k}_2}^{\alpha'_2}(B_2-c\hat{k}_2)$, which shows that it is better for the UAV to only serve hotspot $1$.

If $\alpha'_1\geq \alpha'_2$, we can see that (\ref{equ_proof_compare}) always holds, and the hotspot $1$ with shorter flying distance and larger user demand is superior to hotspot $2$.

%When $k_1^*\neq k_2^*$, note that $R_{k_1^*}^{\alpha'_1}(B_1^0-ck_1^*)\geq R_{k_2^*}^{\alpha'_2}(B_2^0-ck_2^*)$ and $k_1^*, k_2^*$ are the optimal energy allocation, we have
%\bee\begin{split}  \log(\sum_{i=0}^{k_1^*}\frac{1}{i!}(\frac{\alpha'_1 (B_1^0-ck_1^*)}{e})^i)
%\geq&\log(\sum_{i=0}^{k_2^*}\frac{1}{i!}(\frac{\alpha'_2 (B_2^0-ck_2^*)}{e})^i)\\
%\geq &\log(\sum_{i=0}^{k_1^*}\frac{1}{i!}(\frac{\alpha'_2 (B_2^0-ck_1^*)}{e})^i). \end{split}\ene
%Then, similar to the analysis of the case $k_1^*=k_2^*$, we can show that $R_{\hat{k}_1}^{\alpha'_1}(B_2+D_{1,2}-c\hat{k}_1)\geq R_{\hat{k}_2}^{\alpha'_2}(B_2-c\hat{k}_2)$.

For the case $R_{k_2^*}^{\alpha'_2}(B_2^0-ck_2^*)>R_{k_1^*}^{\alpha'_1}(B_1^0-ck_1^*)$, the analysis is similar as above. In this case, actually it is easier to prove hotspot $2$ will keep all energy rather than sharing with hotspot $1$. Therefore, we can conclude that it is better for the UAV to only serve the hotspot with maximum individual expected profit.

\section{Proof of Proposition \ref{pro_forking}}\label{app_pro_forking}

Note that we sort the hotspots by their expected profits received from a single UAV. Thus, if all UAVs choose to jointly serve the same hotspot, they will choose hotspot $1$ and the corresponding expected profit is $\max_{k_{1}}R_{k_{1}}^{\alpha'_1}(B_0-D_1-\frac{ck_{1}}{N})$. If the overall expected profit received from serving both hotspots $1$ and $2$ is larger than $\max_{k_{1}}R_{k_{1}}^{\alpha'_1}(B_0-D_1-\frac{ck_{1}}{N})$, the cooperative UAVs will definitely fork to serve different hotspots. Specifically, the sufficient condition can be written as
\bee\label{equ_thm_forkingmain}\begin{split} &\max_{k_{1}}R_{k_{1}}^{\alpha'_1}(B_0-D_1-\frac{ck_{1}}{N})\\
<&\max_{k_{1}}R_{k_{1}}^{\alpha'_1}(B_0-D_1-\frac{ck_{1}}{N-1})+R_{k_2^*}^{\alpha'_2}(B_0-D_2-ck_2^*). \end{split}\ene
According to (\ref{equ_Rk_expo_go0}), we only need to show
\bee\label{equ_beta12}\begin{split} &\max_{k_1}\sum_{i=0}^{k_{1}}\frac{1}{i!}(\frac{\alpha'_1 (B_0-D_1-\frac{ck_{1}}{N})}{e})^i\\
<&(\max_{k_1}\sum_{i=0}^{k_{1}}\frac{1}{i!}(\frac{\alpha'_1 (B_0-D_1-\frac{ck_{1}}{N-1})}{e})^i)\\
&\times(\sum_{i=0}^{k_{2}^*}\frac{1}{i!}(\frac{\alpha'_2 (B_0-D_2-ck_2^*)}{e})^i). \end{split}\ene

Denote $\beta_1=\frac{\alpha'_2}{\alpha'_1}$. If $\beta_1\in(0,1]$, we have
\bee\label{equ_betain01}\begin{split} &\sum_{i=0}^{k_{2}^*}\frac{1}{i!}(\frac{\alpha'_2 (B_0-D_2-ck_2^*)}{e})^i\\
>&1+\beta_1^{k_{2}^*}\sum_{i=1}^{k_{2}^*}\frac{1}{i!}(\frac{\alpha'_1 (B_0-D_2-ck_2^*)}{e})^i. \end{split}\ene
Therefore, when \bee\label{equ_thm_con1}\begin{split} &\frac{\max_{k_1}\sum_{i=0}^{k_{1}}\frac{1}{i!}(\frac{\alpha'_1 (B_0-D_1-\frac{ck_{1}}{N})}{e})^i}{\max_{k_1}\sum_{i=0}^{k_{1}}\frac{1}{i!}(\frac{\alpha'_1 (B_0-D_1-\frac{ck_{1}}{N-1})}{e})^i}\\
<&1+\beta_1^{k_{2}^*}\sum_{i=1}^{k_{2}^*}\frac{1}{i!}(\frac{\alpha'_1 (B_0-D_2-ck_2^*)}{e})^i, \end{split}\ene
(\ref{equ_thm_forkingmain}) always holds.

If $\beta_1>1$, we have
\bee\label{equ_beta_greater1}\begin{split} &\sum_{i=0}^{k_{2}^*}\frac{1}{i!}(\frac{\alpha'_2 (B_0-D_2-ck_2^*)}{e})^i\\
>&1+\beta_1\sum_{i=1}^{k_{2}^*}\frac{1}{i!}(\frac{\alpha'_1 (B_0-D_2-ck_2^*)}{e})^i. \end{split}\ene
Therefore, when \bee\label{equ_thm_con2}\begin{split} &\frac{\max_{k_1}\sum_{i=0}^{k_{1}}\frac{1}{i!}(\frac{\alpha'_1 (B_0-D_1-\frac{ck_{1}}{N})}{e})^i}{\max_{k_1}\sum_{i=0}^{k_{1}}\frac{1}{i!}(\frac{\alpha'_1 (B_0-D_1-\frac{ck_{1}}{N-1})}{e})^i}\\
<&1+\beta_1\sum_{i=1}^{k_{2}^*}\frac{1}{i!}(\frac{\alpha'_1 (B_0-D_2-ck_2^*)}{e})^i, \end{split}\ene
(\ref{equ_thm_forkingmain}) always holds.

By solving (\ref{equ_thm_con1}) and (\ref{equ_thm_con2}), we obtain the sufficient condition $\frac{\alpha'_2}{\alpha'_1}>\max(\varphi^{\frac{1}{k_2^*}},\varphi)$ as in the proposition.

\bibliographystyle{IEEEtran}

%\bibliographystyle{IEEEtran}
%\bibliography{UPSref}

\end{document}